\documentclass[12pt]{article}

 \newread\testifexists
 \def\GetIfExists #1 {\immediate\openin\testifexists=#1
     \ifeof\testifexists\immediate\closein\testifexists\else
     \immediate\closein\testifexists\input #1\fi}

 \usepackage{gthstyle}\usepackage{amsfonts}
 \usepackage{amssymb}
  \usepackage{graphicx} \usepackage{epstopdf}
 \immediate\openin\testifexists=e
     \ifeof\testifexists\immediate\closein\testifexists\else
     \immediate\closein\testifexists\input e\fipsf

 \def\Bbb#1{\setbox0=\hbox{$\tt #1$}  \copy0\kern-\wd0\kern .1em\copy0}

 \def\bbf#1{\setbox0=\hbox{$#1$} \kern-.025em\copy0\kern-\wd0
         \kern.05em\copy0\kern-\wd0 \kern-.025em\raise.0433em\box0}

 \immediate\openin\testifexists=a
     \ifeof\testifexists\immediate\closein\testifexists\else
     \immediate\closein\testifexists\input a\fimssym.def  

                     \newcommand{\fn}{\footnote}
              
 \newcommand{\be}{\begin{eqnarray}}             \newcommand{\ee}{\end{eqnarray}}
 \newcommand{\bi}[1]{\begin{itemize}\item[#1]}         
       \newcommand{\ei}{\end{itemize}}
 \newcommand{\eqn}[1]{(\ref{#1})}

 \newcommand{\crlb}[1]{\label{#1}\\[2pt]}
 \newcommand{\eela}[1]{\quad\hbox{\scriptsize{#1}}\label{#1}\end{eqnarray}}
 \newcommand{\eelb}[1]{\label{#1}\end{eqnarray}}
 
 \newcommand{\newsecb}[2]{\section{#1}\label{#2}\setcounter{equation}{0}}
 
 \newcommand{\nolabels} {\def\eel{\eelb} \def\crl{\crlb} \def\newsecl{\newsecb}}

\newcommand\publishversion{\nolabels\setlength{\textheight}{9in}\setlength{\oddsidemargin}{0in}
    \setlength{\textwidth}{6.3in}\setlength{\topmargin}{-0.1in}}

 \def\a{\alpha}         \def\g{\gamma}      
 \def\d{\delta}         
 \def\k{\kappa}      \def\l{\lambda}      \def\m{\mu}
             \def\vv{\varphi}    
 \def\j{\psi}            \def\r{\varrho}

     \def\OO{{\mathcal O}} \def\ZZ{\mathbb{Z}} \def\NN{\mathcal{N}}
 \def\pa{\partial} 
  
 \def\dd{{\rm d}}  \def\bra{\langle}   \def\ket{\rangle}

 \def\fract#1#2{{\textstyle{#1\over#2}}}
 \def\ffract#1#2{\raise .2 em\hbox{$\scriptstyle#1$}\kern-.3em/
                 \kern-.2em\lower .15 em \hbox{$\scriptstyle#2$}}
 
 \def\half{\fract12}  
 
 \def\part#1#2{{\partial#1\over\partial#2}}

\def\lowerheightfig#1#2#3{\(\raise-#1\hbox{\includegraphics[height=#2]{#3}}\)}
\def\lowerwidthfig#1#2#3{\(\raise-#1\hbox{\includegraphics[width=#2]{#3}}\)}

\publishversion
\begin{document} \begin{titlepage}

\title{\normalsize \hfill ITP-UU-11/43  \\ \hfill SPIN-11/34
\vskip 20mm
 \Large\bf How a wave function can collapse without violating Schr\"odinger's equation, and how to
understand Born's rule.}
\author{Gerard 't~Hooft}
\date{\normalsize Institute for Theoretical Physics \\
Utrecht University \\ and
\medskip \\ Spinoza Institute \\ Postbox 80.195 \\ 3508 TD Utrecht, the Netherlands \smallskip \\
e-mail: \tt g.thooft@uu.nl \\ internet: \tt
http://www.phys.uu.nl/\~{}thooft/}

\maketitle

\begin{quotation} \noindent {\large\bf Abstract} \medskip \\
 It is often claimed that the collapse of the wave function and Born's rule to interpret the square of the norm as a
probability, have to be introduced as separate axioms in quantum mechanics besides the Schr\"odinger equation. Here we
show that this is not true in certain models where quantum behavior can be attributed to underlying deterministic
equations. It is argued that indeed the apparent spontaneous collapse of wave functions and Born's rule are features that
strongly point towards determinism underlying quantum mechanics.
\end{quotation}
\vfill \flushleft{December 8, 2011; \ last revision July 20, 2012}

\end{titlepage}

\eject

\newsecl{Introduction}{intro.sec}

The observations reported here were made while the author attended a meeting on the foundations of quantum
mechanics.\fn{The Heinz von Foerster 100 meeting on \emph{Self - Organization and Emergence}, Vienna, 11-13
November 2011.} There it was repeatedly claimed that the collapse of the wave function cannot be reconciled with
Schr\"odinger's equation\cite{BGh}--\cite{Adler}, and therefore has to be introduced as a separate axiom.

An important argument that can be brought up in favor of this position is the following: suppose that a system
starting off in a quantum state \(|A\ket_0\) would, after some time \(t\), lead to a collapsed state
\(|A\ket_t\), while a system starting off as \(|B\ket_0\) would end up in the collapsed state \(|B\ket_t\).
Then, where would the state \(\l\,|A\ket_0+\m\,|B\ket_0\) end up? If this isn't the state
\(\l\,|A\ket_t+\m\,|B\ket_t\), would this not imply a violation of Schr\"odinger's equation? Or does it imply a 
breakdown of the quantum superposition principle?

Not only do we never explicitly observe that Schr\"odinger's equation is violated anywhere, but the claim is also
at odds with models that this author brought up to explain quantum mechanics as being the realization of
statistical features of an underlying deterministic theory\cite{GtHbosons}--\cite{GtHentangled}.

Similar statements are encountered concerning the Born interpretation of the wave function as being a description of
probabilities. The probability of finding a system described by a wave function \(|\j\ket\) to be in a certain
state \(|x\ket\) when a measurement is made, is exactly equal to the square of the norm of the inner product
\(\bra x|\j\ket\). This also appears to be a special, separate axiom. What, after all, do probabilities have
to do with equations such as the Schr\"odinger equation?

Born's rule is sometimes assumed to follow from Gleason's theorem\cite{Gleason}, which deduces the dependence of
probabilities exclusively on the square of the norm, from the absence of any non-trivial alternatives. Indeed, \emph{if} probabilities
do depend on wave functions only, then the Born rule is the only reasonable outcome. But this is no derivation of
quantum logic itself, and, as happens more often with purely mathematical theorems, any clues it might give concerning the nature of 
quantum mechanics itself, are misleading.

It is important to clarify these issues. Once and for all? Some issues will not be completely settled with the
arguments presented here, so that the discussions will doubtlessly continue. In particular, Bell's inequalities
will continue to raise questions, but this author is convinced of the basic correctness of the presentation given
below. As for the apparent breakdown of the superposition principle as mentioned above, the answer is simple: in a
deterministic theory, a system can be in a state \(A\) or in a state \(B\), but never in a superposition. More precisely, 
the \emph{sub-microscopic} degrees of freedom that we suspect to be deterministic, may also serve to fix
the \emph{macroscopic} variables describing outcomes of measurements, in a classical statistical sense, which
could be just a perfect explanation of why an apparent collapse takes place. In the scenario embraced in this paper, 
quantum superposition of sub-microscopic states (the states of the ``hidden variables") never occurs! This will be 
the beginning of our answer to the superposition question raised above.

The degrees of freedom in terms of which we usually describe atoms, molecules, subatomic particles and their 
fields will be referred to as \emph{microscopic} degrees of freedom. It is these that have to be described as superpositions 
of the sub-microscopic states, and in turn, the macroscopic states are superpositions of microscopically defined states.
Perhaps the most accurate way to describe the situation is to say that the states we use to describe atoms, quantum fields, etc.,
serve as \emph{templates}. A particle in the state \(|x\ket\) or in the state \(|p\ket\) or whatever, will nearly
always be a superposition of many of the sub-microscopic states; as such, they evolve \emph{exactly} according to Schr\"odinger 
equations. In contrast, the sub-microscopic states evolve classically. The macroscopic states also evolve classically, but
the details of their evolution laws are far too complicated to follow, which is what we need the microscopic template states
for. The \emph{macroscopic} states, such as people and the indicators of measuring devices, are probabilistic distributions of
the sub-microscopic ``hidden variables". We actually only observe macro-states. Our assumption is illustrated diagrammatically in Fig.~\ref{states.fig}.

\begin{figure}[h] \setcounter{figure}{0}
\begin{center}\includegraphics[width=105 mm]{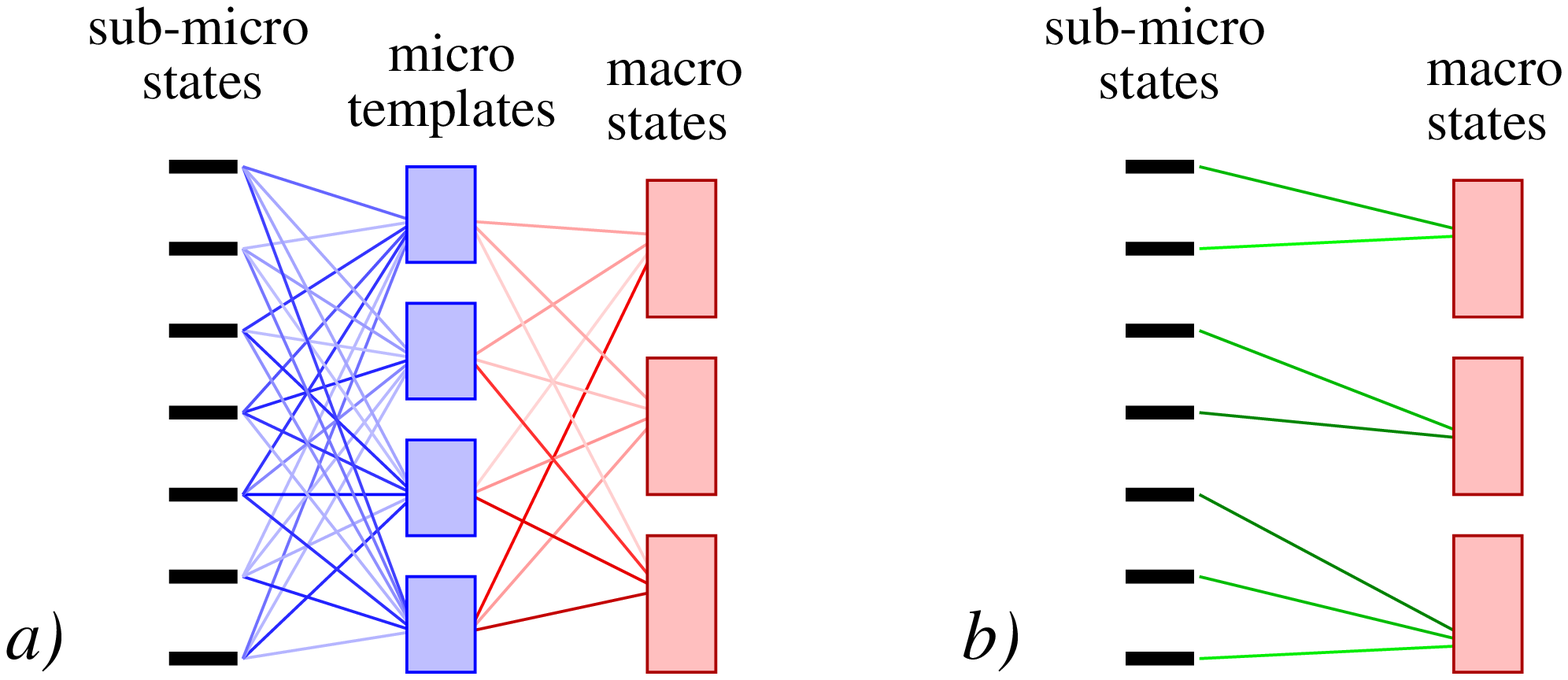} \end{center}           \quad\\[-48pt]
 \begin{quotation}  \begin{caption} 
 {\small Classical and quantum states. $a)$ The sub-microscopic states are the ``hidden variables".  Atoms, molecules and fields are 
 templates, defined as quantum superpositions of the sub-micriscopic states, and used at the conventional microscopic scale. The usual ``classical" objects, such as planets and people, are macroscopic, and again superpositions of the micro-templates. The lines here indicate quantum matrix elements; the second set is essential, as it refers to the environment, see the main text.
 $b)$ The classical, macroscopic states are \emph{probabilistic} distributions of the sub-microscopic states. Here, the lines therefore indicate probabilities.  All states are astronomical in numbers, but the microscopic templates are more numerous than the classical states, while the sub-microscopic states are even more numerous.}
\end{caption}\label{states.fig}\end{quotation}
\end{figure}


We first explain how our suspicion is obtained, in the introductory sections~\ref{determ.sec} and \ref{wave.sec}.
The main argument is presented in terms of the Schr\"odinger cat gedanken experiment (Section~\ref{cat.sec}).
Although a discussion of Bell's inequalities\cite{EPR}\cite{Bell} is not the central theme in this work, they have
to be addressed since most readers will bring forward that the idea of any deterministic model underpinning
quantum mechanics will seriously be affected, if not directly disproved by Bell's observations. The author's view
on this issue is briefly explained in the Discussion Section~\ref{Bell}. It can be summarised as: \emph{if you
believe in determinism, you have to believe it all the way}.  In Section~\ref{note.sec}, we briefly address
criticism that was raised against the first version of this paper. There, also the notion of \emph{superdeterminism} 
is carefully addressed.

\newsecl{Writing deterministic theories quantum mechanically}{determ.sec}

A well-known attempt to formulate a deterministic theory underlying Quantum Mechanics is the De Brogie-Bohm theory of
pilot waves\cite{B1}\cite{B2}. In this theory, the quantum wave function is turned into an ontological object (``pilot wave") that
affects the motion of particles. This behavior can be cast into equations which can be called deterministic, but they carry an essential ingredient that is non-local.
Although the construction of the theory in the present paper is quite different, being more generally applicable to quantum theories, and not necessarily non-local,
there is some similarity in the mathematical sense. The theories are indeed related: in Refs.~\cite{GtHdeterm},\cite{GtHinfoloss}, a model with local information loss is described that
contains a dynamical variable \(\vv(t)\) that may be identified with the phase of a pilot wave. Presently, we will not further discuss the issue of information loss.

In terms of a nomenclature employed by Hardy\cite{hardy}, one could call the de Broglie-Bohm theory \emph{\(\j\)-ontic}, which means that the wave function has an ontological significance. This we regard as a concoction that is actually unnecessary and somewhat misguiding. In the present paper, we shall treat quantum superpositions exactly as in the Copenhagen interpretation; not the wave functions but only the probabilities are physical realities. In Ref.~\cite{hardy}, this is called \emph{\(\j\)-epistemic}. A special case is the ``wave function of the universe", which, in a sense, is ontic; we explain this further in Section~\ref{wave.sec}.

Indeed, many readers will have reasons to doubt our claim that deterministic models can underly the quantum nature of our
world. Yet at least the converse statement, that deterministic models can be described \emph{as if} they were quantum
mechanical, with a wave function that obeys a Schr\"odinger equation, is easy to explain and
justify\cite{GtHbosons}--\cite{GtHentangled}. Consider a deterministic world, for instance one in which a
multi-dimensional continuous real variable \(\vec q(t)\) obeys an equation of the form
 \be {\dd\over\dd t}\vec q(t)=\vec f(\vec q)\ , \eel{deteq}
where \(\vec f(\vec q)\) is any (continuous) vector function of the vector \(\vec q\). One can then introduce
the operator \(\vec p\) by defining
 \be \vec p\equiv -i{\pa\over\pa\vec q}\ , \eel{momdef}
and introduce the Hamilton operator
 \be H=\half\bigg(\vec p\cdot\vec f(\vec q)+\vec f(\vec q)\cdot\vec p\bigg)\ , \eel{Hamdef}
to find that
  \be H=H^\dag\ ,\qquad  \dot{\vec q}=i[ H,\,\vec{q}\, ]\ ;\qquad \dot{\OO\vphantom{|}}(\vec q)=i[ H,\,\OO(\vec q)] \ .
  \eel{queqs}
Consequently, \emph{if} we introduce a wave function \(\j(\vec q,t)=\bra\vec q\,|\j(t)\ket\), then it obeys
the Schr\"odinger equation
 \be {\dd\over\dd t}|\j(t)\ket=-iH|\j(t)\ket\ . \eel{Schroeq}

Also, the norm of the wave function is conserved, because \(H\) is hermitean, so that it can serve to define
probabilities (this was the reason for symmetrizing the expression \eqn{Hamdef}). However, often the Jacobian
\(\vec\nabla\cdot\vec f(\vec q)\) vanishes, and in that case \emph{any} positive function of \(\j(\vec q,t)\)
can serve as a probability, so it is not a priori clear why one should choose the square of the norm.
Nevertheless, our calculations will show that the Born rule is mandatory.

There is an important caveat. Deterministic theories of this sort are not quite the same as quantum theories
because the hamiltonian \eqn{Hamdef} is not bounded from below. It therefore has no ground state and consequently
there is no stability in these theories. One cannot rely on extremum principles in such a situation and after the
dust settles, the resemblance with familiar quantum theories is not quite so big. We stress however that these
should be considered as technicalities. Progress is being made in resolving this issue; this is postponed to
a future publication\fn{Note added in a later revision of this paper: some of these publications now exist 
in preprint form \cite{GtH5}\cite{GtH6}\cite{GtH7}.}. One can imagine, for instance, that some projection operator can be
introduced that singles out a subclass of all states for which the hamiltonian does have a lower bound. The
quantum mechanical procedure is \emph{formally} completely correct. In particular, solving Eq.~\eqn{Schroeq} by
diagonalizing \(H\), requires the introduction of quantum superpositions. What we call quantum mechanics is
nothing but the mathematically convenient procedure of solving Eq.~\eqn{Schroeq} using operators, Fourier
transformations, and so on, leading to the use of particles and fields as templates. This is our starting point.

It is important to realize that theories such as Eq.~\eqn{deteq}, are deterministic only if the initial state
could be defined with infinite precision. The real numbers in the vector \(\vec q\) require the unambiguous
definition of infinite series of digits, otherwise \emph{chaos} generates uncertainties. Therefore, it may be
better to consider a subtle twist to this model by replacing the continuous variables by discrete ones. In that
case, time also has to be chosen discrete. Let now the variables \(\vec q(t)\) be points in a discrete lattice and
let an equation of motion tell how a physical system hops from point to point as a function of a discrete time
variable: \(t\in\ZZ\). Write the equation of motion as
 \be \vec q(t+1)=\vec f(\vec q(t))\ . \eel{discreom}
Usually (but not always) we make the restriction that \(\vec f\) has an inverse. Indeed, a large class of cellular
automaton models can be defined that are time reversible\fn{The author actually suspects that also non-time
reversible cellular automaton models may be considered \cite{GtHdisdet}\cite{BJV}, such as, for example,
\emph{Conway's Game of Life}\cite{conway}, but these are more complicated to handle\cite{GtHbeable}.} and
\emph{universal} by construction\cite{fredkin}. The operation \eqn{discreom} is then simply a permutation, and we
can write it as
 \be |\,\vec q,t+1\ket=\sum_{{\vec q\,}'}U_{\vec q,{\vec q\,}'}\, |\,{\vec q\,}',t\ket\ , \eel{discrevol}
where \(U_{\vec q,{\vec q\,}'}\) is the permutation written as a matrix in Hilbert space, containing only ones
and zeros. As it is unitary, this matrix can be written as
 \be U_{\vec q,{\vec q\,}'}=\Big(\exp(-iH)\Big)_{\vec q,{\vec q\,}'}\ , \eel{UHam}
and we may or may not decide to choose all eigenvalues of \(H\) to lie in the interval \([\,0,\,2\pi)\).

We now  have a more decent hamiltonian since it has a ground state, but we seem to get another
problem: is energy fully conserved or perhaps only \emph{modulo} multiples of \(2\pi\)?  Do the stability problems 
not re-enter through the back door? One would also prefer to have an \emph{extensive} hamiltonian, one that may grow to indefinite values
as we consider larger volumes of space\cite{GtHCA}.

However, one can ensure that the hamiltonian is strictly bounded from below and absolutely conserved, by carefully defining its
eigenstates. Here again, remaining problems can be dismissed as technical ones that have no bearing on the real
issue: we can formally rewrite every time-reversible system (discrete \emph{or} continuous) as if it were a
quantum system with a hermitean hamiltonian, obeying a Schr\"odinger equation. In many cases the resulting hamiltonian 
will show all those features that were thought to be unique to quantum mechanics.

The most important conclusion is that quantum mechanics does not necessarily imply that the underlying theory
is indeterministic. We can use the quantum mechanical machinery to examine deterministic systems as well.

\newsecl{The wave function}{wave.sec}

If a theory is deterministic, we can also choose the initial condition deterministically, that is, there is
exactly one state that is realized at \(t=0\), and it describes ``the universe". Being deterministic, such a
theory describes a single world at all times -- there is never any interference, \emph{in terms of the \(\vec q\) variables}. 
The wave function at \(t=0\) is \(|\j,0\ket\) and, in the continuum case, it could be written as
 \be\bra \vec q\,|\j,0\ket=\NN\,\d(\vec q-\vec q_0)\ ,\qquad\bra \vec q\,|\j,t\ket=\NN\,\d(\vec q-\vec q_t)\ , \eel{initcont}
where \(\NN\) is a normalization constant, while in the discrete case,
 \be\bra \vec q\,|\j,t\ket = \d_{\vec q,\,\vec q_t}\ . \eel{initdiscr}

Of course, such ``wave functions" do not spread. At all times \(t\), only one value of the `ontological'
variable \(\vec q\) is realized. This wave function always takes the form of Eqs.~\eqn{initcont} or
\eqn{initdiscr}. This is why it could be regarded as ontological. Note that, even though it never
spreads nor collapses, it nevertheless fully obeys the Schr\"odinger equation \eqn{Schroeq}.

Now look at our world. At first sight, the wave functions we use to describe it, look very different. But then we
have to realize that we do not know the operators \(\vec q(t)\). The operators that we do use, for instance in
describing the Standard Model, actually refer to states close to the lowest eigenstates of \(H\), so that they
contain low-energy projection operators. Therefore, when expressed in terms of Standard Model (SM) operators, the
observables \(\vec q(t)\) always form non-trivial superpositions of eigenstates of SM operators, probably
compounded by operators of as yet unknown particles and fields at higher energies (the ``hidden" variables).
Thus, the Standard Model describes the real world in terms of templates. We only have access to
a small subset of all templates; those that describe objects beyond the Standard Model, are simply not yet known.

The deterministic, or ``ontological" degrees of freedom may be suspected to describe our world at the Planck
scale. This is why we refer to these as the \emph{sub-microscopic} variables. Atomic, molecular,
and sub-atomic (``Standard Model") degrees of freedom are called `microscopic', while ``classical", visible
degrees of freedom, such as photographic plates, counters and cats, are `macroscopic'.

In terms of the eigenstates of SM operators, the eigenstates of \(\vec q(t)\) must seem to be highly entangled.
This is what happens in some of our models\cite{GtHCA}, and it may well be the reason why, in terms of microscopic
variables, even the wave functions \eqn{initcont} and \eqn{initdiscr} may seem to be complicated entangled ones.

However, common sense gives us the suspicion that the macroscopic observables may be diagonal again in terms of
the \(\vec q(t)\) operators. It is an interesting assumption, which we will adhere to. See Fig.~\ref{states.fig}$b$. It indeed implies that \emph{the wave 
function of the universe} will be \emph{collapsed} in terms of the macroscopic observables. According to our 
deterministic theories, these wave
functions should nevertheless obey the theory's Schr\"odinger equation. Now that we expressed our suspicion that
wave functions can be constructed that always stay collapsed when macroscopic variables are considered, we can
attempt to construct them more directly, starting from conventional theories of physics such as the Standard
Model.

It may be worth-while to emphasize that the assumption that we have an underlying discrete and deterministic theory
is a useful one, but it might be avoided, if so desired. If we do \emph{not} assume the existence of a deterministic explanation 
of quantum mechanics, we can still assume the validity of special classes of wave equations and initial conditions of
the universe that have the same effect: an automatic collapse of the wave function. However, in this case, this effect may
have the appearance of a physically very unlikely conspiracy. Our theory of deterministic, discrete variables (``hidden
variables") is one natural explanation that completely removes these conspiracies, but there might be other such theories.

\newsecl{Schr\"odinger's cat}{cat.sec}

The prototype example is the Schr\"odinger cat gedanken experiment\cite{cat}. Let us start with an over simplified
description that runs as follows. At \(t=0\), we have an unstable atom\fn{Historically, Schr\"odinger thought of an unstable atom, 
but a Stern-Gerlach experiment might be even more illustrative. For our present argument any quantum system can be used.}
in the initial state
 \( |\j(0)\ket=|1\ket\), and a cat, in a sealed box.
A certain moment later, at \(t=t_1\), there is, say, a \(40\%\) chance that the atom decayed into state
\(|2\ket\), by emitting a photon \(\g\). The wave function is then
 \( |\j(t_1)\ket=\sqrt{0.6}|1\ket+\sqrt{0.4}|2,\,\g\ket\).
If the atom at that moment has decayed, the cat is poisoned and dies, so, according to the over simplified argument, we
then have a cat in the super-imposed state \(\sqrt{0.6}\,|\,\hbox{live cat}\ket+\sqrt{0.4}\,|\,\hbox{dead
cat}\ket\). Then, the box is opened for inspection; a ``measurement" takes place.

Will the cat really be in a superimposed state? Of course not. Decoherence takes place\cite{MS},\cite{decoh}---\cite{Zurek}, 
and we expect
that the cat will be \emph{either} in the live \emph{or} in the dead state. The question usually asked is how the
wave function really evolves. Does it ``collapse"?

To do this right, we have to take into account all those physical degrees of freedom that might be responsible
for decoherence. Billions of atoms interact very weakly with the decaying atom and the cat. Each of these atoms can be
in dozens of states, so the total dimensionality of the vector space spanned by these atoms is a huge number,
 \be N=\exp(C\,\k)\ , \eel{astro}
where \(\k\) is the number of atoms and \(C\) a number of order one or larger, so indeed, \(N\) is
astronomically large. As explained in the previous sections, we expect these environment states in general to
be highly entangled, but in the first step of this argument, we consider simple, ``pure" environment states in
their energy eigenstates. These states will be referred to as \(|\,ES,\,t\ket\).

Now, we can consider the above process more carefully. The initial state at \(t=0\) is
 \be |\j,0\ket=|1\ket\,|ES,0\ket\ . \eel{psi0}
Then, at \(t=t_1\), we may assume that the wave function is (apart from an overall phase rotation)
 \be |\j,t_1\ket=\bigg(\sqrt{0.6}\,|1\ket+\sqrt{0.4}\,|2,\,\g\ket\, e^{i\vv(ES)}\bigg)|ES,t_1\ket\ . \eel{psit1}
Here, we take into account that the environment states may have caused a relative phase rotation \(\vv(ES)\).
The point is that total energy is conserved, but a small (positive or negative) part of it may have been
absorbed by the environment, a part that will be \emph{different} when the atom has decayed. Since we are not
closely watching the environment while doing the experiment, we do not have perfect control over this phase
difference.

In Eq.~\eqn{psit1}, for simplicity, the \(ES\) states were considered to be in some eigenstate of the
hamiltonian. But this is not the wave function that we are interested in. Both in terms of the SM degrees of
freedom, and in terms of the ontological states of Section~\ref{wave.sec}, the environment will be time
dependent. Also, the ontological states \(|O,k\ket\) are expected to appear in the conventional quantum
mechanical description as entirely entangled states, which we shall refer to as \(|EES,k\ket\):
 \be |EES,k\ket=\sum_{i=1}^N\a_i^{(k)}|ES_i\ket\ . \eel{entangled}
Let us reexpress the result in terms of a density matrix:
 \be|EES,k\ket\,\bra EES,k|=\sum_{i,j=1}^N\a_i^{(k)}{\a^*}_j^{(k)}\pmatrix{0.6&\sqrt{0.24}\,e^{-i\vv_j}\cr \sqrt{0.24}
 \,e^{i\vv_i}&0.4\, e^{i(\vv_i-\vv_j)}} |ES_i,t_1\ket\bra ES_j,t_1|\ , \eel{density1}
where \(\vv_i\) stands for \(\vv(ES_i)\). Furthermore, \(N\) is the number \eqn{astro} of environment states,
and the index \(i\) labels them.

If all states \(|EES,k\ket\) occur with (approximately) equal probability \(1/N\), we can use orthogonality,
 \be\sum_k\a^{(k)}_i{\a^*}^{(k)}_j=\d_{ij}\ , \eel{orthog}
and the density matrix becomes
 \be\sum_k{1\over N}|EES,k\ket\,\bra EES,k|={1\over N}\pmatrix{0.6\,\mathbb{I}&\sqrt{0.24}\,X\cr\sqrt{0.24}\,X^*&0.4\,
 \mathbb{I}}\ , \eel{ontodensity}
where \(X\) stands for
 \be X=\sum_ie^{-i\vv_i}|ES_i\ket\,\bra ES_i|\ , \eel{Xdef}
 while \(\mathbb I\) is defined to be
 	\be \mathbb{I}=\sum_i |ES_i\ket\,\bra ES_i|\ . \eel{Idef}

  The density matrix is that of a \emph{microcanonical ensemble} where the total energy is fixed, \emph{apart
from small variations that allow our states to depend slowly on time}, so that all states that obey the
restrictions dictated by the macroscopic description of the environment (including the total energy) have
(approximately)  equal probabilities. It is important to realize that this assumes that the $EES$ states that
we use all describe this subset of all macroscopic states. Indeed, this is what our ontological theory
supposes, so there is no contradiction here.

If it weren't for the phases \(\vv_i\), the environment would merely contribute the identity as its density
matrix. But now let us consider the phases in the off-diagonal part. Suppose \emph{decoherence} takes place.
This means that the phases \(\vv_i\) take all values, practically randomly, depending on the energy of the
environment states. These energy eigenstates are delocalized. Therefore, if any matrix element is considered
describing localized operators, many of the $ES$ states contribute, and their contributions are equal apart
from the phases. Therefore, one expects the phases to cancel out. In short, \emph{if used only in combination
of localized observables or operators}, the off diagonal terms in our density matrix, the matrices \(X\),
cancel out to zero. The density matrix is therefore
 \be \r={1\over N}\pmatrix{0.6&0\cr 0&0.4}\mathbb{I}\ . \eel{diagdensitym}

Note that this density matrix was arrived at by performing probabilistic averages, not by demanding a wave
function to collapse. It therefore completely agrees with the Schr\"odinger equation. But now it is of interest to
see what it means in terms of the ontological states \(|O,k\ket\). We stated that the probability for starting out
with any of these states was equal: \(P_k=1/N\). Suppose now that, at the start, we had one single pure state
\(|O,k_0\ket\). The suspicion that we explained in Section~\ref{wave.sec} is that this ontological state will
\emph{either} lead to a dead cat \emph{or} a live one, but never to a superposition. This agrees with our density matrix
\eqn{diagdensitym} if indeed the \emph{probability} that state \(|1\ket\) was realized was 60\% and the
probability for \(|2,\g\ket\) was 40\%. The probability for a superposition to arise is zero.\fn{Note, that the
important assumption mentioned in Section~\ref{wave.sec} was used here. It was assumed that, in the 
deterministic theory, the question whether the cat
is dead or alive can be settled by careful statistical analysis of the state of the sub-microscopic degrees of
freedom of the system. If the wave function is delta-peaked on one of the sub-microscopic states, it will be
delta-peaked as either a live cat or a dead cat.}

Throughout the process, the probabilities for any of the ontological states to be realized were conserved.
Therefore, we must conclude that, from the very start, the ontological states consisted for 60\% of states that
would later evolve into a live cat and 40\% of states that would evolve into a dead one. If indeed we had picked a
state at random, the 60/40 distribution would be that of the probabilities.

This, we now claim, is the origin of Born's rule. The ontological states only evolve \emph{either} into
pure states only describing a live cat, \emph{or} into pure states only containing a dead cat, and never a
superposition. The probabilities are simply in the number of ontological states with these properties. If one
starts out picking one at random, then the probabilities will always be given by Born's rule. Notice that this
argument identifies the Born rule probabilities with the relative abundances of the initial states that could
have been picked ``at random".

In deterministic physics, such as in the classical Van der Waals gas, the origin of probabilistic distributions
can only be in the arbitrariness of the initial state (assuming infinitely precise equations of motion). If one
assumes any kind of continuous distribution of positions and momenta of molecules at \(t=0\), then this determines
the fate of the system, again in probabilistic terms. According to our ontological theory of quantum mechanics,
the probabilities generated by Born's rule, are to be interpreted exactly in the same terms. If we do not know the
initial state with infinite accuracy then we won't be able to predict the final state any better than that. The probabilistic
distribution at \(t=0\) determines the probabilistic distribution at all later times.

An important issue was raised as a possible criticism of the above argument: is it not circular? In a sense, it is. 
We used the Born rule to identify the numbers 0.6 and 0.4 in Eq.~\eqn{ontodensity} as being probabilities. Of course, we
had no choice there. Upon rotating vectors in Hilbert space, only the sums of the norm-squared of the matrix elements
have the appropriate additive conservation properties, including the demand of being non-negative\cite{Gleason}. This part of the 
argument could be called circular. It is the second part that is of crucial relevance: if we side-step the microscopic 
template states and accept the assertion that the \emph{macroscopic} states must be directly linked to the \emph{sub-microscopic} states, 
without quantum superpositions, then the only interpretation left for these same numbers is that they represent relative 
abundances in the initial states. This is then what the Born rule ultimately boils down to.

Let us emphasize again the significance of regarding atoms, photons, spins, as well as quantum fields, as \emph{templates}.
Regardless whether we use the values of fields or the Fock space of particles to describe a basis of states, and
regardless whether these particles are in position space or in momentum space, the total set of these basis elements is much smaller
than the total set of sub-microscopic states. The templates are quantum superpositions of sub-microscopic states. As soon as
we attempt to use \emph{locality} to reduce the known laws of physics in terms of local laws, one encounters all possible quantum 
superpositions of templates and one is tempted to believe that all these superimposed states are physical realities.
The truth is, however, that the sub-microscopic states themselves never undergo quantum superposition. Therefore, one
is not allowed to describe a new physical reality by superimposing ontic states locally, without reference to nearby states
(and possibly also far away states), as if they were not affected by the superposition. This, we now believe, is at the core of the apparent 
contradictions one encounters when working with Bell's inequalities.

It is only if we restrict ourselves to the usual templates of states containing only limited numbers of localized particles,
that we are fooled into believing that the wave function suddenly collapsed when the cat's door was opened, since
we are confronted with the density matrix~\eqn{diagdensitym}; in reality
the ontic states of the underlying automaton correspond to templates that are highly entangled with the environment, 
so that the density matrix~\eqn{diagdensitym}
emerges naturally, fully in accordance with the Schr\"odinger equation.
\newsecl{Discussion. Bell's inequalities}{Bell}

The collapse of the wave function has been approached from many angles in the past (see for instance \cite{Zurek}), and of course we
have Everett's view based on parallel worlds\cite{Everett}, which did receive a certain amount of support in the community. 
In particular the latter view differs from the approach advocated in this paper as to what is
epistemic and what is ontic in our description of Nature.

Our use of the notion of decoherence is admittedly a simplified one. There are many ways to approach this phenomenon\cite{MS}.
What was used in this paper is a residual interaction between the observed system and its environment. Any exchange of
energy causes phase shifts, the angles \(\vv_i\) in Eq.~\eqn{density1}. Our use of decoherence boils down to averaging over these angles \(\vv_i\),
which, as is well-known, leads to diagonal density matrices in the energy frame.

The number \(N\) of Eq.~\eqn{astro} stands for the \emph{dimensionality} of the Hilbert space of environment
states, and for the \emph{total number} of allowed ontological states of the cellular automaton. In the latter
terminology, superpositions are forbidden, so if states \(|A\ket\) and \(|B\ket\) are ontological states, then
\(|\j\ket=\l|A\ket+\m|B\ket\), with \(\l\ne 0\) and \(\m\ne 0\), is not such a state. This is how the states
\(|\hbox{live cat}\ket\) and \(|\hbox{dead cat}\ket\) could emerge as ontological states, but not the state
\(\l|\hbox{live cat}\ket+\m|\hbox{dead cat}\ket\). Thus, the initial state automatically collapses with the appropriate
probabilities.

In our deterministic theory for QM, quantum superposition must be looked at as a property of the statistical
approach to handling the extremely complex local equations of motion. Quantum wave functions were introduced
for the convenience of the computation; linearity came as a handy tool for making calculations, but it so
happens that quantum superpositions of ontological states themselves do not describe any real world, and this,
as it turns out now, explains why we do not see quantum superpositions occurring in the macro world. By using
linearity of the Schr\"odinger equation, we automatically adopt the Born interpretation of the squared norms
as probabilities, because only this way the linear evolution equation for the density matrix \eqn{ontodensity}
can assure probability conservation.

One may even conclude that the absence of superimposed states in the macroscopic world, which is usually
mistaken to imply a collapsing wave function, is actually an important argument \emph{in favor} of microscopic
hidden variables.

However, we do have to face the Bell inequalities.  These inequalities refer to gedanken experiments on states in
which quantum mechanical objects are produced in some quantum entangled states, such as two photons in a state
with total spin zero, and many possible variations of this theme\cite{Seevinck}.

As has been shown with abundant evidence, such states can indeed be produced in real experiments. Observers,
separated by macroscopic distances away from one another, can choose which component(s) of the wave function to
detect, and they can use ``free will" to determine their choices. Bell's inequalities appear to imply that the
correlations then found cannot possibly be reconciled with a deterministic hidden variable theory. In the hidden
variable theories that one then has in mind, the quantum particles are, somehow, accompanied by classical hidden
variables that decide ahead of time what the outcome of any of the possible measurements will be.

Clearly, Bell has shown that such hidden variable theories are unrealistic. We must conclude that our cellular
automaton theory cannot be of this particular type. Yet, we had a classical system, and we claim that it
reproduces quantum mechanics, with probabilities generated by the squared norm of wave functions.

Even though we work with wave functions that are quite complicated quantum superpositions of the SM eigenstates,
we also emphasized that quantum superpositions of ``ontological states" themselves are not ontological. Thus, if
in a Bell experiment one axis was chosen, say, for the measurement of a spin, any other axis where the spin
variable would not commute with the previous one, is in principle forbidden. This would clearly avoid the Bell
inequalities, but of course it raises an important question: how can it be that the gedanken experiments just
described could be performed so easily, not only in our imagination but also in real experiments\cite{aspect}?

The first notion that will have to be scrutinized is the concept of ``free will"\cite{CK}\cite{Newscientist}.
Clearly, an observer who uses ``free will" to choose the direction of a spin to be measured (or any other, more
general basis for a quantum measurement), actually makes his or her decision depending of the outcome of Nature's
laws in his/her own system. This is inevitable in deterministic theories, so we do not have to worry about ``free
will" itself\cite{GtHfreewill}. Yet we still have a puzzle. Suppose, for instance, that the observer's decision is
made to depend on fluctuations of the light from a distant quasar. If Alice and Bob participate in the experiment,
they both use quasars which are located diametrically opposite to one another. These quasars would 
both be spacelike separated
from the device that produced an entangled particle. According to the arguments of the previous sections, the
initial state of the entire system does not allow for quantum superpositions of these ontological states, and this
leads to the apparently inevitable conclusion that the initial object of the quantum measurement in question must
be entangled with the quasar in a very delicate sense, even if they are spacelike separated.

Spacelike correlations can of course be understood if systems have a common distant past; therefore, by itself,
the finding that there must be strong spacelike entanglement constitutes no contradiction, but it seems quite
amazing. What must be kept in mind is that, in our cellular automaton models, the state commonly referred to as
the \emph{vacuum} is by no means featureless; there are vacuum fluctuations, and this means that the vacuum is a
highly complex solution of local field equations. Thus, the vacuum may show all sorts of correlations with matter,
even over large (spacelike) distances.

The fact that the vacuum is far from featureless is exhibited most dramatically by the Hawking
effect\cite{hawking}, the fact that the state perceived as a vacuum by one observer, may seem to be teeming with
particles when viewed by a strongly accelerated observer.

Let us also repeat a remark made in earlier publications: it is clear that the phenomenon of decoherence plays an important role
in our arguments concerning Nature's wave function. If systems without decoherence would exist, they could clash
with our argument. The idea of a quantum computer is based on the assumption that, somehow, decoherence can be
avoided. We now claim that, in a theory of sub-microscopic determinism, decoherence is inevitable, even though it
is totally in accordance with Schr\"odinger's equation; the physical world simply isn't perfectly clean, there are
always disturbances due to interactions.\fn{Indeed, it seems to be harder to describe free massive
particles in terms of deterministic variables than complicated interacting systems using ontological cellular automata.} 
This leads us to a prediction: quantum computers will eventually suffer from decoherence in such a way that they will never
outperform a classical computer, \emph{if this classical computer would be scaled to Planckian dimensions}, that
is, its processors would have switching frequencies of the order of \(10^{44}\) Hertz and memory units containing
one bit of information in every \(10^{-98} \hbox{ cm}^3\) (or perhaps \(10^{-65}\hbox{ cm}^2\)). Of course, real
classical computers never perform that well, so quantum computers will still be interesting technological
challenges, but if it is claimed that they can solve $NP$ problems, such as the factorization of large numbers into
their prime factors\cite{Shor}, our prediction is not a trivial one. The basis
for the prediction, of course, is the fact that, according to this theory, all equations of motion are classical
somewhere near the Planck scale.

\medskip

The author thanks R.~Plaga for lengthy discussions on the Bell inequalities.

\newsecl{Note added}{note.sec}
It was brought to the author's attention that this work was being criticized, and even ridiculed, in several  
weblogs\cite{Motl}\cite{Leifer}\cite{Mallah}. Since these contain many incorrect descriptions of what we are trying to explain here,
resulting as it seems from inaccurate reading\fn{One 
blogger states that my theory is ``massively non-local", whereas non locality is nowhere required in what was stated here. Our escape from
Bell's inequalities is in the use of pure quantum states as templates.}, it is impossible to 
reply to all of them. Particularly when the attitude one encounters becomes dogmatic: ``Bell's inequalities, Hardy's ``paradox", GHZM states, 
Kochen-Specker theorem, free will theorem, and other results uniformly show that the natural phenomena we observe 
have features that can't be compatible with any theory of the type that Gerard 't Hooft is discussing." And so on.

Yet let me make an attempt\fn{besides adding clarifying remarks at various spots in the paper.}. One of the standard 
arguments against theories of the sort discussed here, was put in the form of the Stern-Gerlach experiment. 
The answer here of course is that the states considered in this experiment (and all others of this type) are still microscopic states. 
As such, they are all templates,
not describing the ontological aspects of the real world. Only when finally a measurement is made, the numerous degrees 
of freedom of the environment generate a density matrix that becomes diagonal, without violating Schr\"odinger's equation.  Indeed, the density 
matrix is a handy tool to show how our quantum templates relate to the macroscopic world, which in turn gives us a glimpse of how
the templates are related to the sub-microscopic states. 

The outcome of the experiment is
correctly given by the Born rule, which means that there is a statistical distribution of the expected outcomes, exactly as described in
this paper. The particles with spins up,
down, or sideways do not describe what really goes on, but are templates, complex quantum superpositions of the real
ontological states of the underlying deterministic automaton. The real ontological states cannot be described at all at the level of atoms,
since they include all vacuum fluctuations.

Also, the assertion that the models described in this paper do not even remotely resemble the real world is baseless. We have 
not even started to produce accurate descriptions of cellular automata that match the real world; we actually expect this to require a long route, 
littered with Nobel prizes long before the finish is reached. In the mean time, all we say about this is that the models we should consider will
mostly be discrete and deterministic, which is all that really matters for the argument. More importantly, we \emph{can} construct many 
deterministic toy models\fn{The harmonic oscillator is there, any \emph{finite} dimensional ``Hilbert space", and non-interacting massless fermions,
just to name a few. Producing larger sets of toy models, matching quantized field theories, is a challenge presently under consideration.} that do exhibit genuine quantum behavior at the microscopic level, not suffering at all from any Bell inequality.
These models are indeed instructive to figure out where many of the so-called no-go theorems go astray.

Once we dealt with the Bell inequalities by
introducing all conventional quantum states as templates, we can even ignore the claims that deterministic models have to be non-local.
At the atomic level, the template states accurately obey the quantum Schr\"odinger equation; no deviation of quantum mechanics is
needed at all there, so all experiments that confirm the quantum mechanical departure from Bell's inequalities are accounted for.

The notion of ``super determinism" is often misunderstood. If super determinism were to stand for a kind of pre-determinism, by which
a system re-arranges its initial states in such a way that the final state takes the form desired by some pre-conceived theory, then this would
indeed be too silly to accept. However, in relation to ``free will", a bit more care is called for. If we allow an observer to have the ``free will" to 
project an object against a template \(\bra \l\j_1+\m\j_2|\) rather than the templates \(\bra \j_1|\) or \(\bra \j_2|\), then this does contradict the 
starting point that Nature's ontological states do not allow this, without any \emph{other} changes in the surrounding (``environment") states. 
So, if an observer changes his mind, this change of mind must be ascribed to features of the 
cellular automaton way back in the past. His change of mind must therefore also affect the initial state. There is no
escape to that. This needs not at all be regarded as some diabolic conspiracy, but just a matter of fact: you can't change the course of events
without making a change of the initial states. Any deterministic theory has this feature.

Finally, it was not the main purpose of this paper to enter into this discussion at all, but to argue that the apparent collapse of the wave function
does not at all require departures from a local Schr\"odinger equation, and the Born rule is a natural consequence of the use of quantum
statistical techniques, particularly if an underlying non quantum mechanical theory is suspected.

\end{document}